# CADS: CORE-AWARE DYNAMIC SCHEDULER FOR MULTICORE MEMORY CONTROLLERS

Eduardo Olmedo Sanchez[1] , Xian He Sun[2]

[1]Technical University of Madrid, Madrid, Spain
[2]Illinois Institute of Technology, Chicago, USA

## ABSTRACT

*Memory controller scheduling is crucial in multicore processors, where DRAM bandwidth is shared. Since increased number of requests from multiple cores of processors becomes a source of bottleneck, scheduling the requests efficiently is necessary to utilize all the computing power these processors offer. However, current multicore processors are using traditional memory controllers, which are designed for single-core processors. They are unable to adapt to changing characteristics of memory workloads that run simultaneously on multiple cores. Existing schedulers may disrupt locality and bank parallelism among data requests coming from different cores. Hence, novel memory controllers that consider and adapt to the memory access characteristics, and share memory resources efficiently and fairly are necessary. We introduce Core-Aware Dynamic Scheduler (CADS) for multicore memory controller. CADS uses Reinforcement Learning (RL) to alter its scheduling strategy dynamically at runtime. Our scheduler utilizes locality among data requests from multiple cores and exploits parallelism in accessing multiple banks of DRAM. CADS is also able to share the DRAM while guaranteeing fairness to all cores accessing memory. Using CADS policy, we achieve 20% better cycles per instruction (CPI) in running memory intensive and compute intensive PARSEC parallel benchmarks simultaneously, and 16% better CPI with SPEC 2006 benchmarks.*

## 1. INTRODUCTION

With the advances of multicore architectures, parallel computing has become main stream computing. Multicore processors are fast becoming building blocks in top high performance computing (HPC) systems. While multicore processors are becoming pervasive, many challenges that we did not see with single core processors are coming up. One of the main challenges in utilizing the power of multicore processors is the sharing of resources by multiple cores. Since more than one core try to access the shared resources simultaneously, it is possible that some of them, such as last level cache, memory controller, main memory, memory bus, etc., become a source of contention and poor performance. This challenge is more visible for data intensive workloads, which request for large amounts of data. Such workloads are common in various fields of science. Among the shared resources, memory controller is a major source of bottleneck due to poor memory performance [12, 36]. Since the controller is typically integrated on chip [18, 19], developers have no control over which core has to be given priority, when multiple cores request for memory access simultaneously. Consequently, it is necessary to design efficient memory controller schedulers that manage what requests are allowed to access.

The design of scheduling policy for memory controller has a significant importance in the overall efficiency of the DRAM memory system [20, 28, 29]. The goal of the memory controller in a CMP is to satisfy data access requirements of diverse workloads in a way to improve performance while giving fair priority to memory requests of each core. The workloads can be parallel or sequential, data intensive or compute intensive. They may have regular data accesses that exhibit some locality or random data accesses. Since it is common to run this variety of workloads on multicore systems, which are critical components of HPC machines and data centers, optimized memory controller scheduling has a significant impact. An optimum controller should consider its scheduling decisions based on different memory characteristics of each workload, such as memory intensiveness, origin of memory requests [4] or workload interferences [5]. For example due to the existing gap between the performance of memory and that of the CPUs [12], a compute intensive application can do more computing if their requests are scheduled prior to those of memory intensive workloads [17]. Hence, when memory contention occurs, it is fair not to give the same priority to data requests from various types of applications. It is important to strike a balance in prioritizing multicore memory requests.

We propose a scheduler that is aware of diversity among data requests. We present the design and implementation of an innovative policy, called Core-Aware Dynamic Scheduler (CADS). CADS is based on two key ideas. (1) CADS considers diverse inherent characteristics of the workloads in making its scheduling decisions, making it core-aware (2) CADS dynamically adapts its scheduling policy in real time, optimizing it for the characteristics of currently running mixture of the workloads. These two characteristics make CADS versatile in a broad scope of running conditions and workloads.

To achieve its goals, CADS uses Reinforcement Learning (RL) approach. RL is a field of machine learning that studies how an agent learns to take an action in an environment so as to maximize long-term reward. This technique is able to adapt dynamically to the current environment conditions through interaction with the environment. Ipek et al. [27] proved that using RL in memory scheduling is practical with a preliminary memory scheduler and [34] used RL for self-tuning of multi-tier web systems. While the usage of machine learning in performance optimizations [27, 32, 33] and Ganapathi et al. [31] inspired us, we have overcome various complexity limitations that RL technique poses, and introduced utilizing core-aware characteristics of memory requests to take advantage of locality and bank parallelism. We present a new learning mechanism that solves the scalability problem found in Ipek's work. This makes our scheduler much more conservative with the resources and possible to use in systems with a high number of cores. These enhancements make our scheduling policy more suitable for CMP architectures.

Main contributions of this paper are:

- We discuss the importance of making scheduling decisions in a CMP environment by considering inherent characteristics of the workloads, such as current memory contention, number of accesses or origin of data requests.

- We introduce a novel memory controller, specially designed for CMP architectures, that adapts its scheduling policy in real time depending on the current running workloads.

- We describe the design and hardware implementation requirements of CADS in a memory controller. We compare the performance of the proposed scheduler with other commonly used memory controllers. We evaluated the performance of CADS in running various combinations of PARSEC parallel benchmarks [25] and SPEC CPU 2006 benchmarks [26]. We enhanced M5 simulator [23] to simulate 4, 8, and 16 core processor architectures for our experimental testing. Our experiments indicate up to 20% CPI improvement over conventional FCFS and FR-FCFS schedulers.

The rest of the paper is organized as follows. In Section 2, we motivate our work by analyzing conventional scheduling algorithms and presenting the need for a dynamic scheduling policy. In Section 3, we provide brief background to memory controllers and RL method. CADS structure and its learning mechanism are presented in Section 4. Hardware implementation of CADS is described in Section 5. We present our experimental setup and compare performance of CADS-based memory controller with popular memory controllers in Sections 6 and 7, respectively. Finally, we present the related work in Section 8, and discuss conclusions and future work in Section 9.

## 2. MOTIVATION

Existing memory controller scheduling policies are not effective for CMPs. There are at least two main problems with them. First, many existing memory controller scheduling algorithms use simple policies [1, 2, 3, 13] to increase the throughput blindly without considering characteristics of running workloads, such as their diverse access localities. The most widely used policy in current processors based on this principle is First Ready – First Come First Serve, which first reorders the accesses to memory with the lowest latency and with the goal of increasing bandwidth. These types of policies work successfully for single-core processors. However for CMPs with the possibility of several workloads running simultaneously, they cause an unfair access to the memory by prioritizing workloads that generate accesses with lower latencies (higher row hit rate) and starving workloads with lower row hit rate [5]. Hence, CMPs require core-aware policies to prioritize accesses to take advantage of locality among accesses of each workload and not only the final throughput.

Second, the advanced controllers proposed in research literature are aimed for optimizing a specific characteristic of the workloads or the memories, such as different locality between cores [4], matching a pre-established history of commands [21], or the possibility of parallelism between accesses to different banks [17]. These controllers are based on exploiting a particular characteristic of the memories or the workloads, but do not combine the effect of multiple characteristics. Consequently they are rigid in practicality and they work well only for certain sets of applications [27]. Considering a broad range of workloads that can run simultaneously in a CMP, we need to design more general and flexible memory controllers that can adapt its policy in real time depending on the changing conditions of the workloads to obtain its maximum computer power. To address these two exposed problems we propose CADS.

Our goal in this study is designing a memory scheduler that can share the off-chip DRAM bandwidth efficiently and fairly among the cores. For this, CADS makes its scheduling decisions based on multiple characteristics of the workloads simultaneously, producing a core aware policy that also can adapt its policy depending on the mixture of the current running workloads.

To achieve this goal we use a machine learning technique, Reinforcement Learning. Due to the simplicity of the RL framework in integrating several characteristics simultaneously in making decisions and due to its learning capabilities for adapting the policy at real time, the choice of RL is reasonable.

## 3. BACKGROUND

We briefly discuss the relevant details about DRAM memory systems and memory controllers. We also provide a brief introduction of Reinforcement Learning and the learning mechanism applied in CADS. More details about DRAM scheduling can be found in [1, 6] and about RL in [27, 7, 8].

## 3.1 SDRAM ORGANIZATION AND MEMORY CONTROLLERS

A modern memory system in CMPs consists of one or more DRAM chips composed by banks. Each bank is organized into a two-dimensional array of DRAM cells, rows by columns, and contains a row buffer where the data is written. Each DRAM request experiences different latency depending on the state of the row buffer [1]. Latency is the lowest when the target row is in the row buffer because only a read or write command is needed; this case is called row hit. When the target row is not in the row buffer, i.e. row miss occurs and the latency increases depending if the row buffer is empty or it contains a different row that needs to be written back.

In addition, modern memories allow parallelism among banks. Depending on the burst length parameter of the DRAM, several columns of the same row can be accessed at the same time. These features increase throughput and reduce memory access latency.

CMPs include techniques to allow multiple cores to keep executing instructions without stalling once for each access [9, 10, 11]. Consequently there are several accesses waiting to be processed, that are stored in a request buffer. It is the task of the memory controller to decide what accesses should be served first.

## 3.2 REINFORCEMENT LEARNING AND CADS LEARNING TECHNIQUE

RL is a branch of machine learning that studies how an agent placed in an environment learns how to behave successfully. The goal of the agent is to maximize a defined long-term reward taking an optimum action a at environment state s, where the states determined by a tuple of meaningful features from the environment. The agent-environment interacts as follows: Every time the agent performs an action the environment changes its state, which changes the tuple of features. A reward is then generated to indicate the desirability of the taken action.

The agent should adapt its policy to the changing conditions. To achieve this, the agent uses the rewards as reinforcement. There exist several learning techniques in generating a reward mechanism. For example, Ipek et al. use a table that stores long-term reward associated to every state [15]. On the contrary we define a mathematical model for every action the agent can take in CADS. The mathematical model approximates future reward of taking an action a in state s [8, 14]. Hence, to choose the proper actions we compute the reward at every step and we choose an action that has maximum reward. The model is adapted with the obtained reward from the environment with a widely used RL learning rule Q-Learning [15]. We discuss the model and reward structure in the following section.

# 4.  CADS: CORE-AWARE DYNAMIC SCHEDULING

## 4.1 CADS STRUCTURE

Figure 1 shows the structure of CADS with the RL implementation. As the figure illustrates, the agent is the memory controller and the environment states are defined by a tuple, which consists of DRAM and workload features ($f_0$, $f_1$, …, $f_n$). The possible actions, (action K) considered for the agent to take are to increase the priority of the requests of a core. Consequently, there is a model for each core.

The optimal memory controller policy is determined by the core-models. These core-models are a parametric ($\theta_0,\ldots,\theta_n$) linear function of the DRAM and workloads tuple features determines the environment states. Every environment-feature is taken into account for scheduling decisions, making a core-aware policy. As explained before, the core models adapt at every step with the environmental reward and the current and previous states with the Q-Learning rule, which makes the model dynamically adaptive to the current conditions.

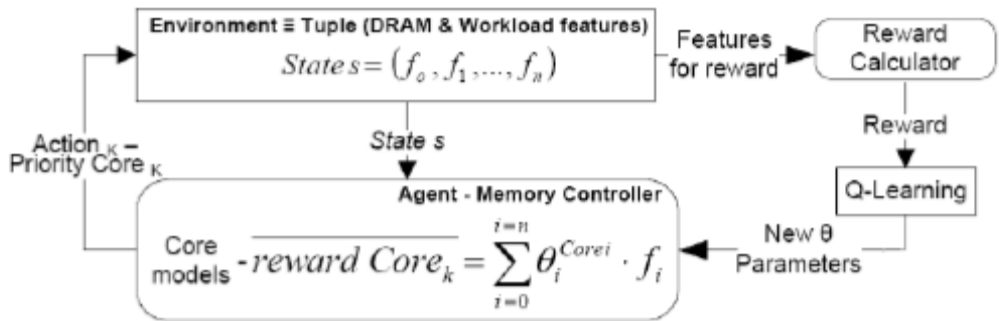

Figure 1. CADS structure

The number of environment-features and possible actions should achieve a good tradeoff between complexity and scheduling performance. Including more features and actions achieve better scheduling decisions [8]; however, the learning time will be longer and it will require more memory and computational resources [27]. Hence, we have limited the number of features and actions as low as possible, while still achieving a good performance with small learning time.

## 4.2 DEFINITION OF THE ELEMENTS OF CADS

In the following subsections, we describe the environmental features that define the state, actions and action models and the reward structure Environment features to define the state:

The features from DRAM and workloads have been selected with the criteria for exploiting only those factors that have a significant performance impact in sharing the memory controller fairly and efficiently.

We classify the features into two types: Local features that depend only on the origin core and Global features that are the same for the whole system. Following are the considered features:

1) Number of petitions waiting to be processed from coren (local feature, abbreviate as NumPet). It is important to differentiate states according to the memory traffic that each workload causes.

2) Number of petitions that will cause a row hit from coren (local feature, RowHitPet). This feature represents the data locality of a workload and is chosen to help the controller learn to differentiate scheduling decisions according to the row locality of each workload.

3) Number of petitions that can be carried in parallel due to bank parallelism (global feature, BPPet). This feature represents the degree of bank parallelism among currently running workloads. It is intended to help the controller to learn how to distribute accesses among different banks.

4)  Number of requests in the recent history of DRAM cycles from coren (local feature, HistPet). This feature helps the controller to differentiate memory intensive workloads from compute intensive workloads, and make scheduling decisions attending to the intensiveness of each workload.

Actions and Action model: The main action for the agent-memory controller is increasing the priority of the requests of a core with memory ready petitions. A request is defined as ready when it can be issued without violating any timing constraints of memory system. If there are multiple accesses from the same core waiting to be processed, then they are issued using the FR-FCFS rule. The action model is used to obtain long-term reward for increasing the priority of accesses from core $C_i$. Our model for each core is based on the parametric linear function using all the features mentioned above.

$$\overline{reward}\ C_i(s) = \theta_0^{Ci} \cdot NumPet + \theta_1^{Ci} \cdot RowHitPet + \theta_2^{Ci} \cdot BPPet + \theta_3^{Ci} \cdot HistPet \qquad (1)$$

Our goal is to modify the $\theta$ parameters according to the state of the system, such as the predicted reward's being as close as possible to the real reward in that state. The global features are the same for each core and the local ones are different. By using independent local features per model (i.e. per core), our scheduler considers different requirements of each workload running in the cores. For this purpose, we use a common RL learning algorithm, called Q-Learning, which we discuss in the next subsection.

**Reward calculator:** The reward calculator generates a reward to modify the core models in a way that the scheduling policy is efficient and fair among the workloads according to the current running conditions. Our reward structure is based on a metric for each core, called memory related starvation (MRStarvation). This metric is defined as the relation between the number of accesses and their stall time:

$$MRStarvation[Core_i] = \frac{Stall\,Time\,[Core_i]}{Number\,Of\,Accesses\,[Core_i]} \qquad (2)$$

In defining the MRStarvation, we consider the number of accesses to differentiate starvation of nonmemory intensive workloads from memory intensive workloads. For example, the starvation value of a non-memory intensive workload is different from that of a memory intensive workload although the stall times are the same because the stall time cycles of the non-memory intensive are more harmful [17]. Consequently, the memory related starvation needs to be larger for the non-memory intensive workload. We define the reward as a function of the current MRStarvation of the cores:

$$reward = f\big(MRStarvation\,[Core_i]\big) \qquad (3)$$

In defining the reward function, we consider the relation between MRStarvations of all cores. Intuitively, if all the MRStarvation values of cores are equal (i.e. our scheduling policy is fair), then the reward should have a positive value. If the MRStarvation values of cores are different (i.e. the scheduling policy is unfair), the reward should be zero. In order to reduce the computation complexity in calculating the variance of memory related starvation values, we obtain our reward as a set of rules between the MRStarvation values. These rules have the following form.

$$if\ (MRStarvation[Core_0] \geq K_0)\&...\&\ if\ (MRStarvation[Core_k] \geq K_3)\ then\ Current\ reward = r \quad (4)$$

Overall, we have 16 different rules in CADS that generate 16 different reward values, each rule compares the MRStarvation of each core with a threshold (K0...K3). Using our observation of various workloads, we classified the possible starvation in four different values: very high, high, low, and very low ($K_0...K_3$). The rules are based on combinations among these four different values, e.g. (very high & high & low & very low) means a very high unfairness, and the rest of the 16 rules are generated in a similar way. We simulated different memory intensive and non-intensive workloads to obtain the minimum and maximum possible values of the MRStarvation cores. Then we divided the resultant interval into four parts, where each part belongs to one of the four values mentioned above.

### 4.3 DYNAMIC ADAPTION OF THE CORE MODELS

To adapt the models to current running conditions, we use Q-Learning [8, 7, 14] algorithm, which is a common update rule for RL. Q-Learning updates the θ parameters of the core models according to the obtained reward, the current state, and the previous state. According to Q-Learning, the rule for updating the θ parameters is:

$$\theta_i^{coren} = \theta_i^{coren} + \alpha \cdot \left[ reward + \gamma \max(Current\ rewards) - previous\ reward \right] \cdot \theta_i^{coren} \quad (5)$$

The parameters are updated when an action a is executed in state s. The max operator means that we need to get the maximum reward of the possible actions that we can take in the state scurrent. There are three constants that we need to define α and ε (0,1], which represent the learning rate and help in convergence of the true rewards [14, 15]. And gamma (γ), where γ is used to dismiss future rewards [8]. The values we use for CADS are α = 0,15, ε = 0,1 and γ = 0,9.

### 4.4 THE CADS ALGORITHM

Figure 3 shows the CADS algorithm. CK[θ0...θ3] and Reward values are initialized to 0 (line 1-2). For every DRAM cycle, we obtain the environment features (line 4), reward of the last memory reordering to update the actions models (line 5), and cores with memory ready petitions that are eligible to be reordered (line 6). CADS policy then chooses the next core for reordering. Using the obtained features, CADS computes the action models by retrieving the core with memory ready petitions with maximum reward (line 7-8). We then choose a core with probability ε based on the FRFCFS rule or the core with the maximum reward value (lines 9-13). The reasoning behind making a scheduling decision using FR-FCFS with a small probability ε [15] is to ensure that the memory controller needs to explore the environment to find the new best actions. There is a possibility of the core models being obsolete due to the dynamic behavior of the workloads. Hence, occasionally, the CADS policy need to choose an action different from the one indicated by the models to adapt to the changing workloads. In these cases, the CADS policy makes a decision using FR-FCFS, since FRFCFS has been shown to be the best policy on average [1].

Once we have selected the core, we reorder the accesses (line 14). The reordering of the requests is done over the baseline of FR-FCFS. CADS gives more priority to those accesses that result in a row hit from the selected core. Among those that are a row hit from the selected core, we give more priority to the oldest request. In case, the selected core by CADS does not cause a row hit and there is a row hit request from a different core, we give more priority to that other request due to its lower access latency. While the petitions are processed by the memory system, we update the action model of the previous reordering using equation (5), with the current maximum reward value, the reward value, and the reward of the previous reordering (line 15-16).

## 5. HARDWARE IMPLEMENTATION

In this section we briefly describe the hardware necessary for implementing CADS. Figure 4 shows the main components. CADS uses two parallel levels to increase performance. In level 1, CADS identifies the core for reordering; and in level 2, it updates the core models. CADS selects the next core for reordering every DRAM cycle and it uses the CPU clock that is several times faster than the DRAM clock to ensure the correct operation. For their operations CADS uses 16 bits fixed point arithmetic and some of the operations are pipelined to adapt the resources to different memory controllers with different timing and size constraints.

---

Algorithm 1: CADS algorithm
1: **Initialize** $C_k[\theta_0...\theta_9]$ to 0
2: **Initialize** prevReward ← 0
3: **for** all DRAM cycles
4:   features ← get_features()  *//Get the current features of the DRAM system*
5:   CurrentReward ← get_reward()  *//Get the reward of the last memory reordering*
6:   CorePetitions ← get_CoresMemReady()  *//Get the cores with memory ready petitions*
7:   CoreMaxReward = getCoreMaxReward(CorePetitions, features)  *//Get the core with the maximum reward*
8:   MaxReward = getReward(CoreMaxReward)  *//Get the reward of the core which reward is maximum*
9:   **if** (rand() < ε) **then**  *//Choose a core based on FR-FCFS with an small ε - probability*
10:   CoreSelected ← select_core_FR_FCFS(CorePetitions)
11:   **else**
12:   CoreSelected ← CoreMaxReward
13:   **end if**
14:   reorder_request_banks(CoreSelected)  *//Reorder the banks attending to the CoreSelected*
15:   update_model(MaxRewardValue, preRewardValue, CurrentReward)  *//Update the model using equation (1)*
16:   prevRewardValue ← MaxRewardValue  *//Store the reward of the current reordering for the next update*
17: **end for**

---

Figure 2. Core Aware Dynamic Scheduling,Algorithm

In the following, we describe the elements that compose the hardware implementation:

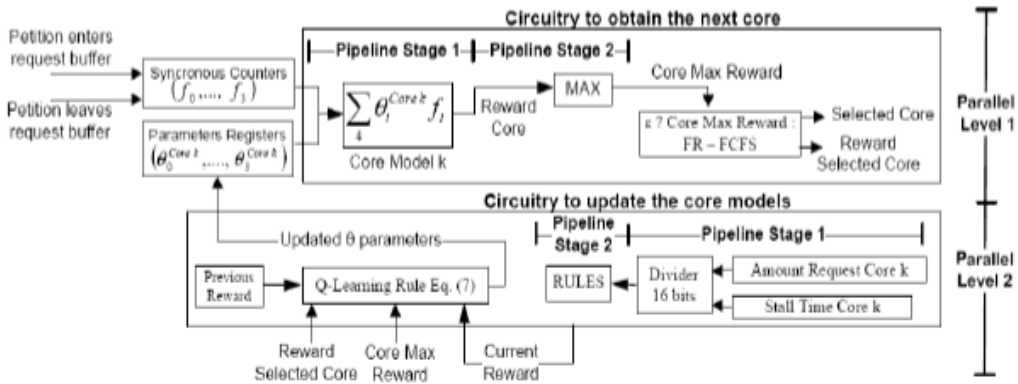

Figure 3. Hardware Implementation

1. Circuitry to obtain the next core: For obtaining the next core for reordering, we first need to compute the core models with the environment features, namely NumPet, BPPet, RowHitPet and HistPet. The features are obtained by incrementing and decrementing synchronous counters when a new request enters or leaves the request buffer. For RowHitPet, an additional register for every core is needed to store the last accessed row. In our experimental analysis, we observe that having the last hundred accesses to the memory (HistPet feature) gives a good performance while keeping the implementation overhead low.

The computation of the core models and the obtention of the maximum can be pipelined in 2 stages to increase speed and to reduce computational requirement. Considering a 3,6 GHz system with a DDR2, 800 MHz memory, the CPU clock is 9 times faster than the memory. We can adjust the pipeline to use four 16 bit multipliers (one per parameter) and four adders per core (one per feature), which needs one cycle computation per core model. As a result of this, the selection of the next core takes a total of six cycles; one cycle to fill the pipe, plus four cycles to compute the models, and another to select between FR FCFS or the selected core.

2. Circuitry to update the action models. The core models are updated with the information of the previous reordering, with the Q Learning rule as shown in equation (7). For updating, we need the reward of the selected core (RewardSelectedCore obtained in stage 1), the maximum reward of the possible actions (CoreMaxReward obtained in stage 2) and the current environment reward. As we explained in Section 4.2, the environment reward is obtained as a set of rules that depend on the values of the cores MRStarvation. The stall time of cores is obtained by using a synchronous counter per core that is incremented every DRA M cycle when there is a request from that core waiting to be processed. The counter is set to zero when all the accesses from that core have been issued. The number of requests coincides with the NumPet feature. The memory related starvation is computed us ing a 16 bit fixed point format divider. This circuitry is pipelined in two stages and for the example considered before, we can compute with a single divider to divide the environment reward in seven cycles: one cycle to fill the pipe, four cycles to compute the MRStarvation for each core, one cycle to select the right rule among the sixteen possibilities, and another one to compute the rule and to obtain the final reward. Since the update of the core parameter models ( $\theta$ $0$ $\theta$ $3$ ) is not in the critical path of selecting the next core we can update the core models in several DRAM cycles if needed, Consequently for the Q Learning rule we can adjust the circuit to use only one multiplier and one adder. Tables 1, 2 detail the memory and computational resources.

Table 1. Memory requirements for a system with K cores, n size Request Buffer, B Banks. And CADS configuration with $\theta$ parameters, f features

| Function | Amount/Size bits | | Function | Amount/Size bits |
|---|---|---|---|---|
| *Syncronous Counter for NumPet* | 1, $\log_2(n)$ | | *Core Model Parameters Register* | 1,16b |
| *Syncronous Counter for BPPet* | 1, $\log_2(n)$ | | *Q-Learning Alpha Register* | 1,16b |
| *Syncronous Counter for HistPet* | 1, $\log_2(100)$ | | *Q-Learning Gamma Register* | *f*, 16b |
| *Syncronous Counter for RowHitPet* | 1, $\log_2(n)$ | | *Computation Model Multipliers* | K·$\theta$, 16b |
| *Syncronous Counter for StallTime* | K, $\log_2(n)$ | | *Computation Model Adders* | 1 |
| *Last Accessed Row hitPet register* | K·$\theta$, 16b | | *Reward Calculator Divider* | 1 |

Table 2. Memory requirements for a system with K = 16 cores, n = 64 size Request Buffer, B = 32 Banks.
And CADS configuration with θ = 4 parameters, f = 4 features

| Resource | Amount/Size bits |
|---|---|
| *Syncronous Counters* | 20, 6 bits & 1, 7 bits |
| *Registers* | 1136 bits |
| *Adders* | 64 |
| *Multipliers* | 4 |
| *Divisor* | 1 |

As stated above, the number of computational resources, such as multipliers, can be adjusted by changing the width of the pipeline. The wider the pipeline i s, the more resources are required and less computational cycles are needed according to the timing and die size constraints.

# 6. EXPERIMENTAL SETUP

We use the M5 Simulator [23] for evaluating CMP architecture with integrated memory controller. M5 is a computer architecture research simulator that provides an out of order CPU that forces execution and timing accuracy supporting varying latencies. For the simulation of the memory system, we have integrated M5 with DRAMsim [24]. DRAMsim is designed for the de tailed simulation of main memory systems and provide all the parameters that can be found in the current memory devices. M5 models an Alpha architecture with enough detail to boot a Linux 2.6 Kernel; the number of cores that can be simulated can be configured from 1 to 64. Different elements that compose a computer system such as caches or buses can be fully configurable by choosing values, such as speed of the bus, size of the cache, cache associativity, etc. M5 simulator offers a fully configurable computer architecture environment that models with enough detailed a computer system that can run Linux kernels. To avoid bottleneck in memory bandwidth as we increase the number of cores, we scale the number of channels in the memory. Such scaling has been use d by Mutlu et al.[17]. The parameters of the simulated multicore CPU are shown in Table 3.

Table 3. Simulation parameters of multicore CPU and memory subsystem

| Component | Parameters |
|---|---|
| Processor | 2.66GHz Alpha architecture, Number of Cores: 4, 8, 16 |
| L1 I-cache | 32 KB per core, 4-way set associative, 64 bytes cache line, 2 latency cycles |
| L1 D-cache | 32 KB per core, 4-way set associative, 64 bytes cache line, 4 latency cycles |
| L2 Shared cache | 2 Mb, 8-way set associative, 64 bytes cache line, 12 latency cycles |
| FSB | 64 bit, 800MHz (DDR) |
| Main Memory | 2GB DDR3 1333-6-6-6, 64 bit, burst length 8, $t_{cas}$=12ns, $t_{faw}$=40, $t_{ras}$=40, $t_{rc}$=54, $t_{rcd}$=12, $t_{rfc}$=280, $t_{rrd}$=8, $t_{rp}$=12, $t_{dqs}$=2, $t_{rwtr}$=12 [30] |
| Channel/Rank/Bank | 2/2/8 (a total 32 banks), Number of channels scaled with the cores 1, 2, 4 channels / 4, 8, 16 cores |
| SDRAM Row Policy | Open Page |
| DRAM controller | On-chip, 64 size buffer request, delay 10 cycles |
| Memory Access Pool | 32 queues for each bank, each queue size is 16 entries |
| OS | Linux 2.6.14 |

For evaluation, we have used two sets of benchmarks from PARSEC benchmark suite [25] and from SPEC CPU2006 suite [26]. PARSEC is a benchmark suite composed by multithreaded applications specially designed to be a representative of the next generation of parallel workloads that we will run on CMP systems. SPEC CPU2006 benchmarks are widely used benchmarks for testing any component of computer architecture research. It is a suite based on current scientific and engineering applications.

Table 4. Benchmarks selected from PARSEC suite          Table 5. Benchmarks selected for SPEC MCPI

| Benchmark | Memory Type | Rate Requests | Number of Threads |
|-----------|-------------|---------------|-------------------|
| a – Dedup | Intensive | 100 | 4,8,16 |
| b – Canneal | Intensive | 85 | 4,8,16 |
| c – Blackscholes | Non-Intensive | 43 | 4,8,16 |
| d – Streamcluster | Non-Intensive | 44 | 4,8,16 |

| Benchmark | Memory Type | MCPI |
|-----------|-------------|------|
| e – 437.leslie3d | Intensive | 7.30 |
| f – 429.mcf | Intensive | 6.45 |
| g – 471.omnetpp | Non-Intensive | 1.96 |
| h – 464.h264ref | Non-Intensive | 0.48 |

We have selected four benchmarks from each of PARSEC [25] and SPEC CPU2006 [26] benchmark suites. Among these benchmarks, two are memory intensive and two are non memory intensive, from each suite. We define memory intensiveness as the ratio of the number of instructions that are reads and write sting accessing the DRAM and the total number of instructions. The number of threads is scaled with the number of cores. The selection for SPEC CPU2006 benchmarks is based on the definition of intensiveness given by Mullet al. [5, 17], called Memory Cycles per Instruction (MCPI). In Tables 4 and 5, we show the selected benchmarks from PARSEC and from SPEC CPU2006 suites, respectively. The problem size selected for PARSEC is sim medium and for CPU2006 we run for 300 million of instructions in a representative part of the program chosen via simulations with M5. The benchmarks are labeled with an alphabet, from a-h .

## 7. PERFORMANCE EVALUATION AND ANALYSIS

We compare the performance of CADS with two common reordering policies FCFS, FR FCFS We analyze the performance of memory controllers using three metrics: (1) The number of requests, i.e. the total number of petitions (reads and writes) that workloads send while they are being executed. This metric is intended to describe traffic in the memory system. If the request buffer is full, new requests have to be re issued by processor cores, which increases the number of memory requests. Hence, lower number of requests indicates lower contention. (2) The waiting latency is the time that requests in the memory request buffer wait (i.e. time between their arrivals in the request buffer till they are processed by the memory system). The purpose of this metric is to show if the memory scheduling policy is able to reduce waiting time of requests by taking advantage of their locality and parallelism. Again, the lower waiting latency indicates better performance of the memory controller. (3) The CPI is the average number of clock cycles per instruction for a workload. This metric is intended to show impact of the memory scheduling policy on overall performance (execution time) of an application.

Table 6. Sets of Experiments from benchmark suites          Table 7. Sets of Experiments from benchmark suites

| Suite | Category | Benchmark |
|-------|----------|-----------|
| PARSEC | Intensive-1 | ab |
| PARSEC | Non-intensive-1 | cd |
| PARSEC | MN-1 Mix | bc |
| PARSEC | MMNN-1 Mix | abcd |

| Suite | Category | Benchmark |
|-------|----------|-----------|
| CPU2006 | Intensive-2 | ef |
| CPU2006 | Non-intensive-2 | gh |
| CPU2006 | MN-2 Mix | fg |
| CPU2006 | MMNN-2 Mix | efgh |

In Table 6 and 7, we show various sets of experiments, where we combine running various workloads. Since our goal is to show how CADS can adapt to memory access characteristics of workloads, we select a mixture of memory intensive and non intensive benchmarks from both PARSEC and SPEC CPU2006. From PARSEC suite, we run a set of two memory intensive benchmarks (Intensive 1), a set of two non intensive benchmarks (Non intensive 1), a mix of one memory intensive and one non intensive benchmarks (MN 1), and a mix of two memory intensive and two non intensive benchmarks (MMNN 1). We form similar sets of experiments with SPEC CPU2006 suite. The different workloads that compose the experiments start simultaneously and are run until completion. The scheduling of the threads and the assignment of the workloads to the cores is done by the Linux kernel. For both PARSEC and SPEC, we use the same memory and CPU configuration.

## 7.1 ANALYSIS OF THE NUMBER OF MEMORY REQUESTS

Figures 4, 5, and 6 show the number of requests to the DRAM in a 4, 8, 16 CMP, respectively, for PARSEC experiments and CPU2006 experiments. In each experiment, we compare the number of requests to access the DRAM using FCFS, FR FCFS, and CADS memory controller scheduling policies. The figures are normalized to the FCFS policy. As we can see in these figures, FCFS has the highest number of requests for each set of experiments and for every policy. The number of requests for FR FCFS is less than that of FCFS, and CADS has the least number of memory requests.

The reduction in the number of requests with CADS policy is due to a reduction in the contention. When the memory request buffer is full of requests, the new ones that come cannot be stored for scheduling. The processor cores need to issue them again, which increases the number of requests and consequently increasing contention. CADS reduces contention by making a better use of locality and bank parallelism of accesses than FR FCFS. In addition, CADS increases throughput because it is able to reorder the accesses by giving more priority to requests from non intensive workloads that spend more time executing instructions and less waiting in stall cycles.

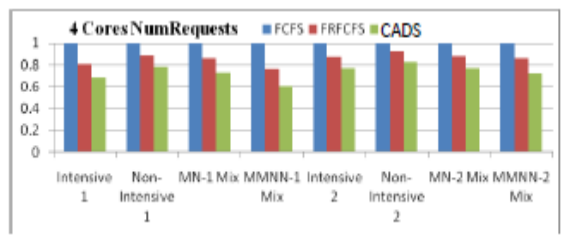

Figure 4. Number of requests for 4 cores.

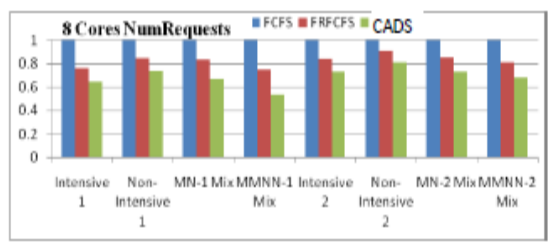

Figure 5. Number of requests for 8 cores

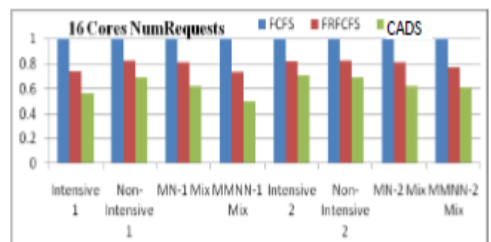

Figure 6. Number of requests for 16 cores

For PARSEC benchmarks, CADS policy outperforms the FR-FCFS policy by 14% on average for Intensive workloads and by 11% on average for Non-intensive workloads. With CPU2006 benchmarks, the average performance improvement with CADS is 11% and 10% for Intensive and Non-Intensive workloads, respectively. This different improvement between the Intensive and the Non-Intensive is because when there is more contention by multiple cores in accessing the DRAM, the performance of FR-FCFS becomes worse. Our CADS policy is able to achieve better performance, producing a higher throughput independent of the contention

For PARSEC benchmarks, the improvement in the non-mixed workloads is 13% and that for the mix is 18%, and for CPU2006 the improvement for the non mix is 11% and for the mixed workloads is 14%. This difference between the mixed and non-mixed is because CADS is specially designed for scheduling in mixed environments, where the workloads are different. FR-FCFS does not differentiate the workloads as opposed to CADS, causing more requests and a bigger unfairness. In addition, in MMNN-1 the average improvement among all the cores is 20 % and in the MN-1 the improvement is 16%. The reason for this is that the higher the contention is the worse the performance is of FR-FCFS. CADS based policy provides better performance in these cases.

As the number of cores in a CMP increases, we can see that CADS has a better performance than the other policies. For PARSEC benchmarks, CADS based policy out performs FR-FCFS by 13%, 15%, and 18% with 4, 8, and 16 cores, respectively. For CPU2006 the performance differences are for 11%, 12%, and 14% with 4, 8, and 16 cores, respectively. As we increase the number of cores, we have more concurrent running threads, resulting in more contention and more mixed workloads. In these conditions CADS can show a better performance than FR-FCFS.

For CPU2006 benchmarks, we observe smaller differences in performance gains with CADS than those for PARSEC benchmarks. Since CPU2006 benchmarks are not parallel benchmarks the contention is smaller. In addition, since it is harder to make use of the bank parallelism techniques using single thread CPU2006 benchmarks, there are fewer petitions coming to different banks concurrently. Hence, as the number of threads increases, as in PARSEC benchmarks, we have more possibilities for improving throughput with bank parallelism.

## 7.2 ANALYSIS OF THE WAITING LATENCY

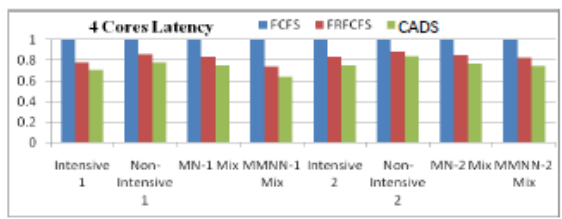

**Figure 7. Latency for 4 cores.**

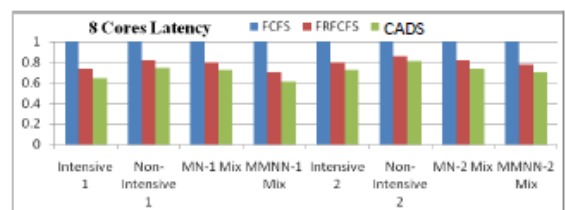

**Figure 8. Latency for 8 cores.**

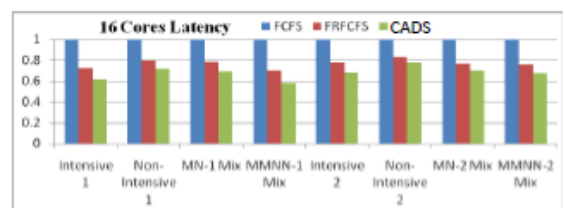

**Figure 9. Latency for 16 cores.**

In Figures 8, 9, and 10 we illustrate the waiting latency for the PARSEC and CPU2006 experiments with 4, 8, and 16 core CMPs, respectively. The figures are normalized to the FCFS waiting latency. From the figures, we can see that there is a significant performance difference between FR-FCFS and FCFS. Since FCFS does not try to reduce the latency by giving more priority to the row hit accesses. Instead, FCFS serves the oldest accesses first causing a worse performance. CADS based policy outperforms FR-FCFS in all the experiments.

For PARSEC benchmarks, CADS is better than FR-FCFS by a 9% on average and for CPU2006 it is better by 7%. CADS policy works well due to the reasons we mentioned earlier: generation of less contention and utilizing bank parallelism. Since there is less contention among memory accesses, they need to wait less time for other accesses and as a result the waiting latency is less. CADS based policy is able to achieve low waiting latency values by scheduling requests that accesses different banks, a factor that FR-FCFS does take into account. As we see from these experiments, we can reduce the latency not only by blindly increasing the row hit rate, but also by reducing the contention and increasing bank parallelism.

The waiting latency with CADS policy is much lower than the conventional policies for both the PARSEC than CPU2006 benchmarks. As we increase the number of cores CADS policy performs even better. As we mentioned earlier, the reason for such improvement is CADS policy's fair scheduling of memory requests from multithread applications, which FR-FCFS does not consider. We also see the same trends of much better performance with CADS policy for experiments where memory intensive and non-intensive benchmarks are mixed.

### 7.3 ANALYSIS OF THE CYCLES PER INSTRUCTION

We compare the cycles per instruction (CPI) values to show the impact of memory controller scheduling on performance of the overall application. We compare the CPI values of the three memory scheduling policies in Figures 11, 12, and 13 for all the PARSEC and CPU2006 experiments, with 4, 8 and 16 core CMPs, respectively. The CPI values reflect the consequences of reducing latency and the contention. The more we reduce these two factors the less CPI we achieve. Consequently, we can see similar performance trends as expected.

FCFS has the worst performance, and is highly outperformed in many cases by FR-FCFS. CADS based policy has the lowest CPI compared to that of FCFS and FR-FCFS. The performance difference between CADS and FR-FCFS is bigger for the mixed workloads. For PARSEC, the average improvement with CADS over FR-FCFS in the mixed workloads is 16 % and that in the non-mixed workloads is 12%.

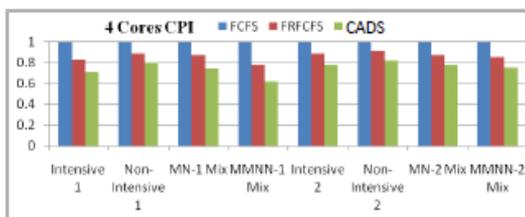

Figure 9. CPI for 4 cores.

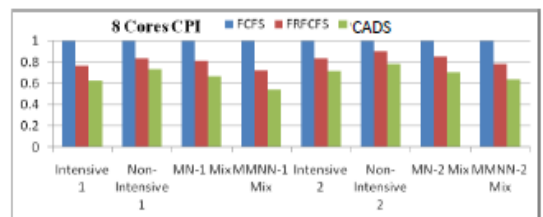

Figure 10. CPI for 8 cores.

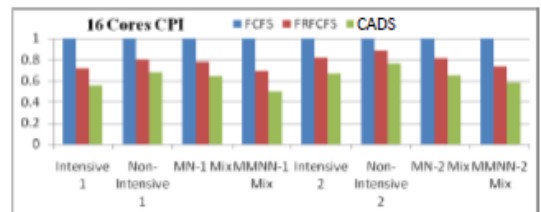

Figure 11. CPI for 16 cores.

As the number of cores increases, the CPI improvements are bigger mainly due to CADS policy's efficient handling of requests from multiple threads. For PARSEC, the average performance improvement with CADS over FR-FCFS is 13% for 4 cores, 14% for 8 cores, and 16% for 16 cores. With CPU2006 it is 10% for 4 cores, 13% for 8 cores, and 14% for 16 cores. The biggest improvement is reached for the MMNN-1 for 16 cores that is 20 %. CADS is able to adapt to the different workloads, which is proven with a significant reduction in CPI values. We have also proven that CADS based memory controller can adapt to various loads at runtime.

## 8. RELATED WORK

Many memory scheduling policies have been proposed to improve the efficiency of memory accesses in the context of single-core processors. Most of these scheduling algorithms use simple policies [1, 2, 3 13] to increase throughput without taking inherent behavior of memory accesses, such as locality of each workload. Rixner et al. [1] proposed a policy called bank-first scheduling scheme, in this policy, memory operations to different banks are allowed to proceed before those to the same bank, thus increasing access concurrency and throughput due to bank parallelism. All of these policies do not adapt to different workloads dynamically, which is one of the key features of our CADS. Some recent policies proposed in research literature [4, 5, 21, 22] are based on specific rules targeted to optimize a specific characteristic of the accesses, loosing the generality that CADS has to be applied for a broad scope of workloads.

There are other memory schedulers that consider the access pattern information for scheduling memory requests. Huret al. [21] proposed an adaptive history based scheduling, that tracks the access pattern and makes its scheduling decisions based on matching a certain pattern of reads and writes. However, the length of the access pattern considered by them is very short. They consider a history of only two commands. Due to this short command history and their implementation of the scheduling policy in a finite state machine, their method is hard to be used in CMP architectures, where the number of possibilities and options on the command history is very large.

Hiroyuki Usui et al. [35] studies the integration between multiple CPU cores and hardware accelerators, common in the current SoC in high performance computing environment, their controller is designed around pre designed ideas that are known to increase performance in general situations, such as deadlines, memory intensive applications , short deadlines. The advantage of this design is that you can obtain a well all around solution for most of the cases.

Fang et al. [4] proposed taking the origin of the workload for scheduling decisions due to better locality of the requests coming from the same source. Core awareness is given focus in this work. Our CADS based scheduling policy is much more sophisticated to include core awareness and adaptively change fairness and bank parallelism much better.

Mutlu et al. in [5] proposed a policy for CMP that considers the fairness of the memory scheduler. On its STFM policy they obtain a metric of the starvation values of the running hardware threads, which is defined as the relation between Tshared (the slowdown experienced by the workload in a share d DRAM system) and Talone (the stall time the thread would have had if it had run alone). STFM takes its scheduling decisions by trying to equalize the thread slowdown between threads and comparing it with a threshold. STFM shows the importance of fairness in CMP, which we have incorporated in our scheduler. The purpose of STFM scheduler was to keep fairness between threads on multiple cores and did not focus on facilitating an application running on multi core processor , t his work is similar to [37 ]], in which a memory scheduler is designed around the idea of keeping fairness by differentiating the tasks of the memory controller into three types and then creating rules around these three types.

Mutlu et al. in [7] introduced PARBS, which focuses on increasing fairness and on using bank parallelism to improve throughput. In PAR BS the accesses are grouped into a batch and the policy ensures that those accesses are served before other batch assuring fairness. For grouping the accesses the authors consider the number of accesses to different banks and the number of prioritize accesses. PAR BS shows an important advance over STFM, and it takes some characteristics of the workload into account. The problem that PAR BS shows is that it cannot learn from its scheduling decisions. On the contrary, in addition to considering fairness and bank parallelism our scheduler is based on mathematical models, which makes it adaptive to the workloads and is much more flexible. Also, using easy integration of features in making scheduling decisions, we think CADS scheduler is expandable.

We are aware of only one work that uses RL for memory scheduling. Ipek et al. [27] proposed an RL technique to schedule the possible commands that a memory controller can issue, such as pre charge row or write data into the memory. Their work is based on increasing the throughput by issuing the commands that are going to provide the biggest throughput. As opposed to CADS, their controller does not differentiate distinct workloads. In addition, since they do not consider the fairness of the current memory policy, their scheduling policy can cause unfairness in issuing memory requests related to different workloads. The hardware implementation overhead of their work is also high since their RL implementation is based on a table, which grows very large if it considers more features and actions. This makes implementing their policy in CMPs impractical as the number of cores increase. In contrast, we replace the table using linear models, which reduces the memory requirements greatly at a low cost of computing the model.

# 9. CONCLUSIONS AND FUTURE WORK

Existing memory controllers in CMPs schedule memory accesses using fixed policies. These policies cannot adapt to the diverse workloads running simultaneously. Trying to increase the throughput of memory accesses without considering the memory characteristics of workloads does not work successfully for CMPs. In this study, we introduced a novel memory controller scheduling policy based on RL, called CADS. T he CADS scheduling policy can dynamically adapt to the workloads that are running simultaneously and can consider the characteristics of the workloads, such as memory intensiveness, locality, and bank parallelism, in making its scheduling decisions. CADS can differentiate scheduling decisions from one workload from another's and provides fairness for workloads in accessing the memory by considering the memory related starvation.

We have shown that CADS successfully reduces the number of accesses and waiting latency in accessing memory for a variety of workloads. As a result, CADS reduces CPI and proves that it can improve overall performance of workloads. Our simulations with M5 and DRAMSim in a 16 core CMP, shows that CADS outperforms FR FCFS by reducing the CPI by 16 % on average (up to 20%) for a mix of memory intensive a non memory intensive PARSEC parallel benchmarks and by 14 % on average (up to 16 %) for CPU2006 benchmarks. We conclude that CADS matches the characteristics that a memory controller for CMP should have, and we believe that the use of the RL technique, on which CADS is based, can be useful for other research areas.

In our future work, we intend to explore the usage of CADS in future many core and heterogeneous architectures, where the cores are specialized in different compute intensive and memory intensive workloads. Due to their specialized core architecture, the workloads running on a core and their requirements can be very different from those running on another core. In this case, it is necessary to take the origin of the requests and their characteristics into account. Since CADS is scalable to consider multiple features, we are planning to integrate different priorities assigned by the OS that the workloads can have into CADS. We believe CADS can achieve further improvements with such extensions and can be better integrated in CMPs.

**AUTHORS**

Eduardo Olmedo Sanchez, graduated from Technical University of Madrid as an engineer in Automation and Electronics researcher in topics related to the application of automation to computer engineering and computer architecture in particular.

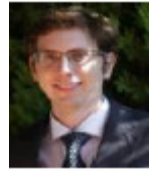

Dr. Xian He Sun is the director of the SCS laboratory in the Illinois Institute of Technology. His current research interests include parallel and distributed processing, memory and I/O systems, software system for Big Data applications, and performance evaluation and optimization.

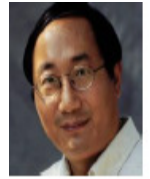